\documentclass[aps,pre,amsmath,amssymb,prereprint,superscriptaddress]{revtex4-2}
\usepackage{float}
\usepackage{graphicx}
\usepackage{xcolor}
\usepackage[export]{adjustbox}
\usepackage[utf8]{inputenc}
\usepackage{bm}
\usepackage{hyperref}
\usepackage{indentfirst}
\usepackage{mciteplus}

\begin{document}

\title{Inertial active particles in a Poiseuille flow: negative mobility and particle separation}

\author{Ankit Gupta} 
\email{ankitgupta@kgpian.iitkgp.ac.in}
\affiliation{Department of Physics, Indian Institute of Technology Kharagpur, Kharagpur 721302, India}

\author{P. S. Burada }
\email{Corresponding author: psburada@phy.iitkgp.ac.in}
\affiliation{Department of Physics, Indian Institute of Technology Kharagpur, Kharagpur 721302, India}

\date{\today}

\begin{abstract}
The diffusive behavior of small entities is strongly influenced by the flow of the surrounding medium, which is ubiquitous in natural and artificial environments.   
In this study, we investigate the transport characteristics of the inertial active Brownian particles (ABPs)
in a microfluidic channel under a Poiseuille flow. 
The interplay between the inertia of the particles and the imposed fluid flow leads to interesting diffusive behaviors.   
For instance, in the overdamped regime ($m \to 0$), particles exhibit a negative average velocity $\langle v \rangle$ due to upstream movement. As $m$ increases, particles tend to move along the flow direction with an increase in $\langle v \rangle$ in the positive direction, exhibiting a maximum at optimal $m$, and diminish for higher $m$ values. 
The effective diffusion coefficient $D_{eff}$ also shows a peak at this optimal $m$. Interestingly, at higher $m$ values, $D_{eff}$ decreases with increasing the noise strength.  
The self-propelled velocity of the particles further enhances the upstream movement. 
Further, the rotation rate of the particles also contributes positively to the upstream motion, and enhances the diffusion of the particles by many orders in the limit of higher $m$. This study reveals that inertia not only modifies swimmer–flow interactions but also enables new dynamical regimes, where mass-dependent trajectories can be harnessed for selective control. Such control holds promise for mass-based particle separation in precisely engineered environments and lab-on-a-chip devices for technological applications.

\end{abstract}

\maketitle
\section{Introduction} 
The Microorganisms or the artificial swimmers, while swimming in a complex fluid under the geometrical constraints, often lead to interesting transport phenomena~\cite{Ramaswamy,Lauga_2009,Marchetti2013,Elgeti2015,Bechinger2016}. The study of such systems has revealed behaviors absent in passive particles, e.g., persistent swimming, accumulation near the boundaries, collective pattern formation, and non-equilibrium phase transitions ~\cite{Bechinger2016,Lauga_2009,Peng2020,Khatri2022}. Unlike passive particles, which move only under external forces, ABPs generate their own propulsion through internal energy conversion, as seen in bacterial suspensions ~\cite{Galajda2007,SchwarzLinek2012,VillalobosConcha2025} and engineered microswimmers, i.e., self-propelled colloids~\cite{Palacci2014,Kesteren2023}.

The collective dynamics of ABPs in shear flow are of fundamental importance across natural and engineered environments, from open oceans to narrow capillaries~\cite{Zoettl2012,Bechinger2016}. Microswimmers such as bacteria, sperm cells, and other microorganisms often navigate through confined spaces under the influence of background flows, as seen in pathogen transport in lung mucus~\cite{Levy2014}, microbial motion through porous media~\cite{Bhattacharjee2019}, sperm motility in the female reproductive tract~\cite{Rutllant2005}, and blood cell transport through the vascular system~\cite{Engstler2007}. Artificial microswimmers—including colloids and microrobots—are likewise designed to function in structured, flowing environments with applications in drug delivery~\cite{Ma2015}, water remediation~\cite{Soler2014,Gao2014}, particle separation~\cite{Reguera2012,Burada_Chemphy}, and disease diagnosis~\cite{Jin2014}. Most studies, however, focus on the low-Reynolds-number regime where inertia is negligible, revealing phenomena such as wall accumulation~\cite{Elgeti2013,Rusconi2014} and upstream swimming in Poiseuille flow~\cite{Rusconi2014,Dey2022}. By contrast, much less is known about the inertial regime, where viscosity and inertia compete, leading to qualitatively new behaviors such as negative mobility and mass-based particle separation.

In recent years, experimental and theoretical studies have shown how inertia affects the unsteady propulsion of ciliated and larger swimmers ~\cite{Wang2012,Ren2017,Scholz2018,Loewen2020,Maity2022}. The recent advancement of high-speed tunable microswimmers \cite{Kim2022,Aghakhani2020} has brought this inertial regime into practical relevance, revealing opportunities for mass-dependent sorting of ABPs in flows. Beyond fundamental interest, particle separation based on mass in controlled environments opens up diverse applications, from selective particle sorting in microfluidics to effective filtration and the design of smart materials.
For instance, in cancer research, the mass of cancer cells can be different from healthy ones, and this difference in mass can be used to identify and isolate cancer cells for further study~\cite{Suresh}. Similarly, mass-based separation methods can detect and isolate specific proteins or pathogens in a sample associated with a specific disease. 
However, separation of entities based on their mass requires a predictive understanding of how particle mass, self-propulsion, and external flow influence each other to shape the particle trajectories in confined boundaries. 
Addressing these issues is the central aim of the present study.
Earlier studies on the diffusion of microswimmers confirm that Poiseuille flows induce trajectory asymmetries, validating flow-driven manipulation approaches. Using the differential interaction of the particles with the flow field, it is possible to achieve separation. Importantly, while the segregation outcome manifests itself in distinct peak positions in particle velocity, these positions are not directly tunable through system parameters alone. Rather, our study highlights how the interplay of inertia and flow geometry establishes natural separation regimes, which can then be exploited for efficient and precise particle filtering in flat channels.

In this article, we study the diffusive behavior of ABPs in a microfluidic channel under the influence of external Poiseuille flow. 
By negating particle-wall interactions, we analyze how the mass of the particles and the self-propulsion speed determine the transport characteristics of ABPs. Previous works have identified swinging and tumbling regimes for rigid active particles in similar flows~\cite{Zoettl2012,Zoettl2013}, but the influence of inertia and mass heterogeneity on these regimes has not been explored. 
Our results reveal distinct migration behaviours tied to inertial effects, suggesting a mechanism of particles for selective transport and separation in microchannels.
This article is organized as follows. 
In Sec.~\ref {sec:model}, we introduce the model system of ABPs driven in a two-dimensional microfluidic channel under an externally imposed Poiseuille flow. 
The transport characteristics of ABPs are presented as a function of various system parameters in Sec.~\ref {sec:transport}. 
The discussion and the main conclusions are presented in Sec~\ref{sec:discussion} and Sec~\ref{sec:conclusions}, respectively.

\section{MODEL}
\label{sec:model}

\begin{figure}[ht]
\centering
\includegraphics[scale=1.15]{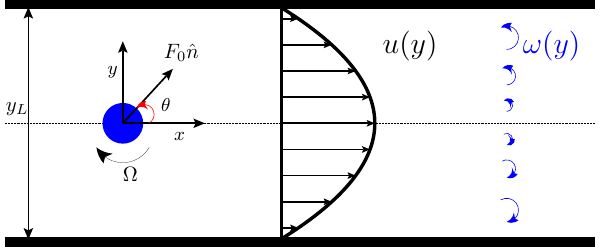}
\caption{
Schematic of an ABP in a 2D microfluidic channel under a Poiseuille flow $u(y)$.
The particle is located at $r = (x, y)$ and has a constant intrinsic speed in the direction of its orientation $\hat{n}$. 
The active force $F_0$, self-propelled angle $\theta$, chirality $\Omega$, local shear rate $\omega(y)$, and width of the channel $y_L$ are indicated.}
\label{fig:channel}
\end{figure}

Consider the dynamics of an ABP of mass $M$ suspended in a thermal bath and constrained to diffuse in a two-dimensional (2D) flat channel subjected to a Poiseuille flow, as shown in Fig.~\ref {fig:channel}.
The particle self-propels in the direction of its orientation with a constant velocity $v_0 \,\hat{n}$, where \( \hat{n} = (\cos \theta, \sin \theta) \) is the unit vector. Here, \( \theta \) is the angle measured relative to the channel axis, i.e., $x$-axis.
The Poiseuille flow \( u(y) \), which is directed along the $x$-axis and has a shear gradient along the $y$-axis with a local shear rate $\omega(y)$. 
Further, ABP may experience external rotation rate (chirality) $\Omega$, where the sign of $\Omega$ defines the handedness.

The equation of motion of ABP is described by the Langevin equations as
\begin{align} 
\label{eq:Langeven-1}
M\frac{d^2\vec{r}}{dt^2} &= - \gamma_t\frac{d\vec{r}}{d t} + F_0 \hat{n} + \gamma_t \, u_s (y) \, \hat{x} + \sqrt{2 D_0} \, \vec{\xi}(t) \\
\gamma_r\frac{d\theta}{d t} & = \omega_s (y) + \Omega + \sqrt{2 D_{\theta}} \, \chi(t)\,,
\label{eq:Langeven-2}
\end{align}
Eq.~\eqref{eq:Langeven-1} and Eq.~\eqref{eq:Langeven-2} govern the translational and rotational motion of ABP, where $\vec{r}$ represents its position vector in 2D. $D_0 = (k_B T /\gamma_t)$ and $D_{\theta} = (k_B T/\gamma_r)$ are the translational and rotational diffusion constants, respectively.
The self-propelled force $F_0 = \gamma_t \,v_0$. Here, $k_B$ is the Boltzmann constant, and $T$ is the temperature of the surrounding medium. $\gamma_t$ and $\gamma_r$ denote the translational and rotational friction coefficients, respectively.
$\vec{\xi}(t)$ and $\chi(t)$ are zero-mean Gaussian white noises, which model the translational and rotational Brownian fluctuations due to the coupling of the particle with the surrounding medium, both obeying the fluctuation-dissipation relation 
$ \langle \xi_i(t)\xi_j(t') \rangle = 2\delta_{ij} \, \delta(t-t')$ and 
$ \langle \chi_i(t)\chi_j(t') \rangle = 2\delta_{ij} \, \delta(t-t')$ for $i, j = x, y$.

A 2D channel is described by two parallel walls with a separation distance \(y_L\), see Fig.~\ref{fig:channel}. 
The Poiseuille flow imposed in this channel is prescribed as
\begin{equation}
    u(y) = u_0 \left[ 1 - \left( \frac{2y}{y_L} \right)^2 \right],
    \label{poe_eq}
\end{equation}
 where \(u_0\) is the flow strength, and the corresponding local shear rate is calculated as
\begin{equation}
    \omega(y) = -\frac{1}{2} \frac{d u(y)}{dy} = \frac{4u_0 y}{y_L^2}\,,
    \label{eq:w_y}
\end{equation} 
which is depicted by the blue arrows in Fig.~\ref{fig:channel}. 
Although the flow velocity decreases away from the centerline and vanishes at the channel walls \(y = \pm y_L/2\), the local shear rate increases linearly with \(y\), see Fig. \ref{fig:channel}. 
At the upper wall (\( y > 0 \)), the local shear rate is in a counter-clockwise direction, and at the lower wall (\( y < 0 \)), the local shear rate is in a clockwise direction. 
Due to the differential rotation arising from the asymmetric shear forces acting on the particles, the particles that are closer to the channel center are dragged more strongly by the flow compared to the particles near the wall.

In order to have a dimensionless description, we scale all length variables by the width of the channel \(y_L\), i.e., \( x \to x/y_L \) and \( y \to y/y_L \). Similarly, we rescale the time as \( t \to u_0 t / y_L \). With this, we can rescale the other parameters as \( D_t \to D_0 / (y_L u_0) \), \( D_r \to y_L D_\theta / u_0 \), and \( \Omega \to y_L \Omega / u_0 \). 
The dimensionless Langevin equation reads, 
\begin{align} 
\label{eq:Lan-dl}
m\frac{d^2\vec{r}}{dt^2} &= - \frac{d\vec{r}}{d t} + f_0 \hat{n} +  u (y) \, \hat{x} + \sqrt{2 D_t} \, \vec{\xi}(t) \\
\label{eq:Lan-dlth}
\frac{d\theta}{d t} & = \omega (y) + \Omega + \sqrt{2 D_r} \, \chi(t)\,,
\end{align}
where the dimensionless mass reads $m = u_0 M/y_L \gamma_t$, $\tau_{m} = M/\eta_t$ is the time scale associated with the inertia of the particle. $f_0 = F_0/ \gamma_t u_0$ is the dimensionless active force. In the rest of the paper, we use dimensionless variables. We assume that ABPs are point-sized entities without interacting with each other. 
By doing so, we can focus solely on the influence of confinement, self-propulsion, and inertial effects on single-particle dynamics, without the added complexity of collective behavior. 
Eqs.~\eqref{eq:Lan-dl}\,\&~\eqref{eq:Lan-dlth} are solved using the standard stochastic Euler algorithm over $2 \times 10^4$ trajectories with sliding boundary conditions at the channel walls \cite{Gupta2023,Ao2014,Khatri2022}.
Since particles along the $y$ direction are confined, we calculate the average velocity $\langle v \rangle$ and the effective diffusion coefficient $D_{eff}$ of the ensemble of ABPs along the channel direction ($x$ direction), in the long-time limit, as 
\begin{align}
\label{eq:meanvel}
\langle v \rangle & = \lim_{t\to\infty} \frac{\langle x(t) \rangle}{t} \,,\\
D_{eff} & = \lim_{t\to\infty} \frac{\langle x^2(t) \rangle - \langle x(t) \rangle^2}{2 \, t}\,.
\end{align}

\section{Transport characteristics}
\label{sec:transport}

In the absence of external forces or strong interaction between the particle and the walls, the steady-state distribution of the particle's position across the channel can be reasonably approximated as uniform. This approximation is valid when the particle's self-propulsion and rotational diffusion lead to a homogeneous exploration of the available space, and the channel width is not extremely narrow compared to the particle's persistence length. Under this assumption, the steady-state probability distribution for the particle position is
\begin{equation}
P(y) = 1, \quad -\frac{y_L}{2} \leq y \leq \frac{y_L}{2}.
\end{equation}
In this case, the average velocity of the particle in the $x$-direction can be obtained as
\begin{align}
\langle v \rangle_{max} = \int_{-\frac{y_L}{2}}^{\frac{y_L}{2}} u(y) P(y) dy.
\end{align}
Thus, the maximum average velocity of the particles in the Poiseuille flow is
\begin{equation}
\label{eq:vx}
\langle v \rangle_{max} = \frac{2u_0}{3y_L}.
\end{equation}
Note that if the particles accumulate near the channel boundaries due to inertia or some interaction, the distribution $P(y)$ would no longer be uniform, and the average velocity could be less than this value.  
Further, this maximum value is the same for both passive and active particles. 

\subsection{Impact of particle mass}

\begin{figure}[ht]
\centering
\includegraphics[scale=1.5]{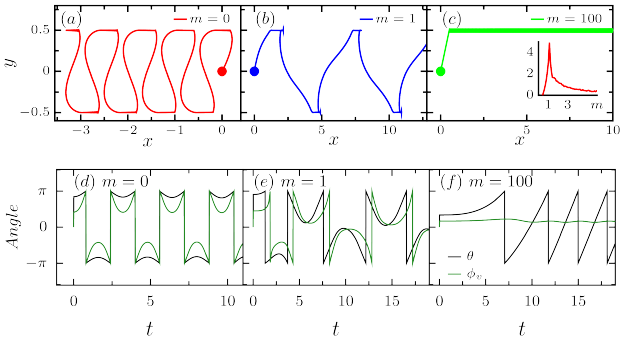}
\caption{(a)-(c) Deterministic trajectories of an active particle in a two-dimensional channel with a Poiseuille flow prescribed by Eq.~\eqref{poe_eq}. The particle starts at the center of the channel, i.e., $x = y = 0$, with different values of particle mass. 
(d)-(f) are the corresponding orientation $\theta$(t) and the instantaneous velocity $\phi_v$(t) directions. 
For $ m\neq 0$, a lag develops between $\theta$(t) and $\phi_v$(t) due to the inertia of the particles. 
Trajectories are obtained by integrating Eq.~\eqref{eq:Lan-dl} and Eq.~\eqref{eq:Lan-dlth} numerically for 
$D_t = D_r = 0$, $\Omega = 0$, and $f_0 = 1$.}
\label{fig:trajectories}
\end{figure}

\begin{figure}[ht]
\centering
\includegraphics[scale=1.2]{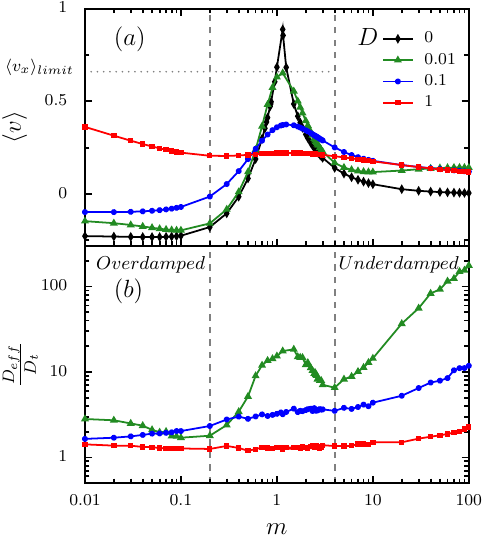}
\caption{
(a) The average velocity $\langle v \rangle$ and (b) the effective diffusion coefficient $D_{eff}$  as a function mass $m$ of the particles for various values of the diffusion coefficient $D=D_r=D_t$. The horizontal dashed line indicates the maximum flow velocity in the channel, $\langle v_x \rangle_{limit}$ (Eq.~\eqref{eq:vx}). The other parameters are set as, $\Omega = 0$ and $f_0 = 1$.}
\label{fig:vel_deff}
\end{figure}

To analyze the effect of the Poiseuille flow on inertial ABP dynamics, in Fig.~\ref{fig:trajectories}, we show the deterministic trajectories of the particle confined in a two-dimensional channel under a Poiseuille flow, 
as described by Eqs.~\ref{eq:Lan-dl} \& \ref{eq:Lan-dlth}. 
In the overdamped regime (see panel (a)), the particle swims upstream the background flow, executing symmetric, channel-spanning oscillations, a behavior driven by the interplay of self-propulsion and flow-induced reorientation. As inertia becomes significant (panel (b)), the particle trajectory transforms into a non-harmonic, asymmetric path, persistently biased downstream. 
The wavelength $\lambda$ of the oscillatory trajectories increases with $m$, reaches a maximum around the optimum mass $m_
{op}$, where the flow strength and mass are nearly equal. 
See the inset of Fig.~\ref{fig:trajectories}(c). 
Further increase in $m$ leads to particle accumulation near the channel boundaries [see inset in Fig.~\ref{fig:trajectories}(c)] due to an increase in the inertia of particles. 
As a result, $\lambda$ decreases. 
In the strongly underdamped regime (panel (c)), the particle ceases cross-channel oscillations entirely and instead self-propels along one of the side walls in a quasi-linear path, revealing a form of inertial wall-locking. 
This leads to a reduction in $\langle v \rangle$ (see Fig.~\ref{fig:vel_deff}).
The panels (d)-(f) in Fig.~\ref{fig:trajectories} depict the corresponding temporal evolution of the orientation of the particle $\theta(t)$ and the angle of its instantaneous velocity direction $\phi_v(t)$. As the mass increases, a growing phase lag emerges between $\theta(t)$ and $\phi_v(t)$, highlighting the delay due to inertia of the particles \cite{Scholz2018} in the reorientation response of the particle to hydrodynamic shear (Eq.~\ref{eq:w_y}). 
Interestingly, these simulated trajectories closely mirror those observed in recent experiments on active Janus colloids and motile microswimmers in confined geometries under flow~\cite{Dey2022}. The emergence of upstream oscillatory motion, downstream drifting states, and wall-bound trajectories dependent on inertia suggests that our model captures essential features of active particle behaviors in microchannels.

The transport characteristics of ABPs are depicted in Fig.~\ref{fig:vel_deff} for different values of $D = D_t = D_r$.
In the overdamped regime ($m \to 0$), the activity of the particles and the local shear rate lead to an upstream drift, evidenced by a negative average velocity $\langle v \rangle$, as shown in Fig.~\ref {fig:vel_deff} and supported by Fig.~\ref{fig:trajectories}(a). 
As $m$ increases, $\langle v \rangle$ shows transitions from negative (upstream) to positive (downstream), marking a crossover point. 
Beyond this, $\langle v \rangle$ increases with $m$, and $\langle v \rangle$ becomes positive, i.e., currents in the positive $x-$ axis of the channel, for different values of noise strength $D$. 
Around the critical value of mass $m\equiv 1$, the system enters an intermediate regime where both inertial and viscous forces are comparably influential. Here, particle inertia introduces a velocity relaxation timescale $\tau_m$, which significantly affects transport behavior. The peak represents the optimal mass where particles can effectively utilize both their self-propelled motion and the favorable flow conditions. 
Since we considered $D = D_t = D_r$, the peak position is independent of $D$. However, if $D_t \neq D_r$, there will be some shift in the peak position with a change in noise strength [see Fig.~\ref{fig:D_plot}(a)\&(c)]. 
Note that for the non-deterministic case, i.e., $D \neq 0$, $\langle v \rangle$ does not exceed the limit $\langle v_x \rangle_{limit}$ (Eq.~\eqref{eq:vx}), which corresponds to the situation where particles are uniformly distributed along the transversal direction. 
On further increase in $m$, there is a gradual decay in $\langle v \rangle$ as the inertia of the particle becomes dominant. 
The effective diffusivity $D_{\mathrm{eff}}$ quantifies channel exploration rather than net drift. 
For $D=0$, $D_{\mathrm{eff}}$ is nearly zero since deterministic trajectories remain confined, and cannot be shown in the log-log plot.

\begin{figure}[ht]
\centering
\includegraphics[width=\linewidth]{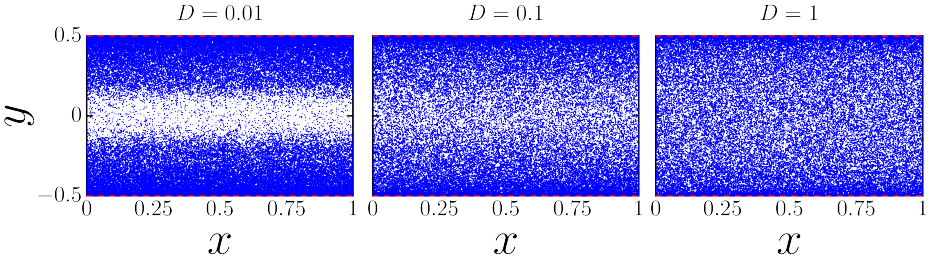}
\caption{
The steady state distribution of chiral particles, for various values of the diffusion coefficient $D$. The other parameters are set as, $m=100$, $u_0=1$, $\Omega = 0$ and $f_0 = 1$.}
\label{fig:snap}
\end{figure}

Introducing a weak noise ($D = 0.01$) produces a pronounced maximum in $D_{\mathrm{eff}}$ at an intermediate mass, where inertia extends orientational persistence while noise destabilizes periodic orbits. At larger noise strengths, dispersion grows more broadly and $D_{\mathrm{eff}}$ increases monotonically with mass. 
In the higher mass regime, $\langle v \rangle$ is nearly independent of the noise strength $D$ due to inertia, and particles exhibit a constant, low  $\langle v \rangle$. 
Interestingly, in this regime, $D_{eff}$ decreases with increasing $D$. 
For lower values of $D$, particles mostly accumulate near the channel boundaries and spread along the length of the channel (see Fig.~\ref{fig:snap}). This leads to higher $D_{eff}$ and lower $\langle v \rangle$. 
With an increase in $D$, particles tend to distribute all over the available space, leading to a uniform distribution along the traversal direction of the channel. Due to the external flow, all the particles tend to move together (with a lower $\langle v \rangle$), which leads to less spreading or lower $D_{eff}$. See Fig.~\ref{fig:snap}. 

\subsection{Influence of translational and rotational diffusion constants}

\begin{figure}[ht]
\centering
\includegraphics{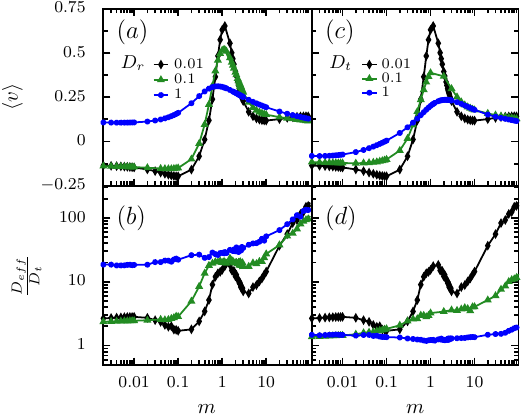}
\caption{(a) The average velocity $\langle v \rangle$ and (b) the effective diffusion coefficient $D_{eff}$  as a function mass $m$ for various values of the rotational diffusion coefficient $D_r$ with $D_t=0.01$. (c) The average velocity $\langle v \rangle$ and (d) the effective diffusion coefficient $D_{eff}$  as a function of mass $m$ for various values of the translational diffusion coefficient $D_t$ with $D_r=0.01$. The other parameters are set as, $\Omega = 0$ and $f_0 = 1$.}
\label{fig:D_plot}
\end{figure}

So far, we consider the case of $D = D_t = D_r$. 
Figure~\ref{fig:D_plot} illustrates the dependence of $\langle v \rangle$ and \( D_{\text{eff}} \) on particle mass \( m \), for different  values of translational diffusion \( D_t \) and rotational diffusion \( D_r \). The observed trends can be attributed to the complex interplay between thermal noise, inertia, and shear-induced alignment. At lower values of \( D_r \), due to the applied Poiseuille flow, local shear rate \( \omega(y) \) aligns particles preferentially against the flow direction, leading to a suppression of \( D_{\text{eff}} \). In this regime, the orientation angle \( \theta \) evolves slowly, enabling particles to maintain alignment upstream for extended durations and thus resulting in a net negative drift. As \( D_r \) increases, particle movement becomes more random, reducing the average alignment and consequently diminishing the upstream drift. This leads to a progressive decrease in the peak height of \( \langle v \rangle \) 
and an increase in $D_{eff}$, particularly for particles with moderate mass, where inertial effects play a role but are not dominant. 

On the other hand, translational diffusion \( D_t \) plays a relatively minor role in the overdamped regime but becomes increasingly significant in the intermediate and underdamped regimes. For moderate mass values, lower \( D_t \) preserves directional motion, resulting in a more pronounced peak in \( \langle v \rangle \), whereas higher \( D_t \) introduces stronger translational noise that disrupts persistent motion and reduces \( \langle v \rangle \). 
As observed before, particles having higher mass, \( \langle v \rangle \) is nearly independent of both $D_t$ and $D_r$. 
However, in this regime, $D_{eff}$ reduces with increasing strength of $D_t$ as discussed before in Fig.~\ref{fig:vel_deff}. 
Collectively, these results demonstrate that while rotational diffusion primarily controls alignment and drift in the (\( m \rightarrow 0 \)) limit, translational diffusion plays a vital role in the intermediate to underdamped regimes.

\subsection{Effect of active force }

\begin{figure}[ht]
\centering
\includegraphics{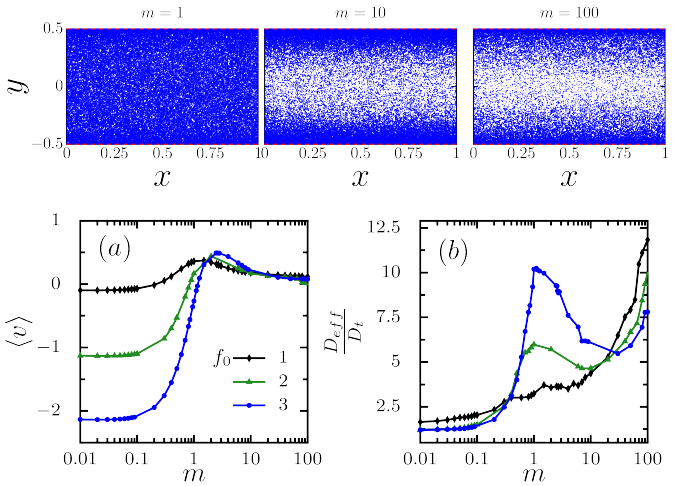}
\caption{
(a) The average velocity $\langle v \rangle$ and (b) the effective diffusion coefficient $D_{eff}$  as a function mass $m$ for various values of active force $f_0$. Additional parameters in the system are set as $D_r$=$D_t=0.1$, and $\Omega = 0$. 
The top panel corresponds to the steady state distribution of active particles, for various values of the mass $m$, for $f_0 = 3$.}
\label{fig:f_plot}
\end{figure}

Figure~\ref{fig:f_plot} $ (a)$ and $(b)$ shows the dependence of \( \langle v \rangle \) and \( D_{\text{eff}} \) on the particle mass \( m \) for various values of the self-propelled force \( f_0 \) respectively. It is evident that $f_0$ significantly modulates \( \langle v \rangle \) especially in the overdamped regime. In this limit, increasing \( f_0 \) enhances the particle's ability to swim upstream against the background flow, resulting in enhanced \( \langle v \rangle \) in the negative $x-$ direction.  
Since particles move together \( |\langle v \rangle| \) is high and \( D_{\text{eff}} \) is low. 
As $m$ increases, inertial effects begin to dominate, and the efficiency of upstream swimming is reduced. 
As observed before, around \( m = 1 \), \( \langle v \rangle \) becomes positive, indicating a transition to a downstream-dominated transport regime. The corresponding $D_{eff}$ is increased monotonically, and the magnitude is greatly enhanced with the increase in $f_0$.  
However, for higher $f_0$, e.g., $f_0 = 3$, $D_{eff}$ shows a non-montonic behavior for higher values of $m$. 
The latter can be understood with the help of the steady state distribution of ABPs for different values of $m$ as depicted in the top panel of Fig.~\ref{fig:f_plot}.
For (\( m > 1 \)), the particles become increasingly entrained by the flow. As a result, \( \langle v \rangle \) decays and $D_{eff}$ slows down. In this regime, as observed before, particles start accumulating near the channel boundaries and spread along the channel length.  
In the limit of higher $m$, due to inertia \( \langle v \rangle \) approaches zero. 
However, $D_{eff}$ increases because of the noise effects and external flow.

\subsection{Effect of external rotation rate}

\begin{figure}[ht]
\centering
\includegraphics{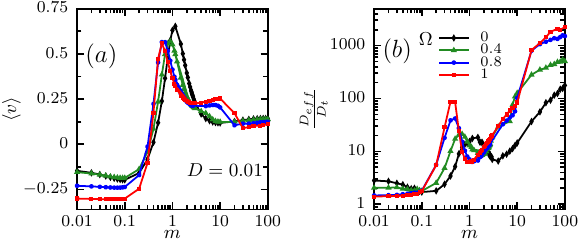}
\caption{
(a) The average velocity $\langle v \rangle$ and (b) the effective diffusion coefficient $D_{eff}$ as a function of mass $m$ for various values of the rotation rate $\Omega$. Additional parameters in the system are set as $D_r = D_t = 0.01$ and $f_0 = 1$.}
\label{fig:omega_plot}
\end{figure}

Figure~\ref{fig:omega_plot} $(a)$ and $(b)$ illustrates the behavior of \( \langle v \rangle \) and \( D_{\text{eff}} \) on \( m \) for various values of the rotation rate \( \Omega \) of the particles. 
The behaviors reveal that \( \Omega \) significantly influences the dynamics of the particles.
In the $m \to 0$ limit, at higher values of \( \Omega \), particles exhibit a stronger upstream drift, as reflected by increasingly negative values of \( \langle v \rangle \). This enhanced upstream motion indicates a more pronounced rotational activity that drives particles against the flow direction. Moreover, the peaks in both \( \langle v \rangle \) and \( D_{\text{eff}} \) -which correspond to optimal drift and diffusion—shift toward lower mass values as \( \Omega \) increases. This shift can be understood as a balance between persistence and inertia. When orientation changes rapidly at large $\Omega$, only lighter particles with shorter inertial relaxation times can adapt effectively, shifting the optimum toward lower mass. For intermediate masses, where inertial and rotational timescales are comparable, inertia suppresses rapid reorientations without acceleration. 
This partial alignment between self-propulsion and rotational drift yields a modest enhancement in \( \langle v \rangle \) and \( D_{\text{eff}} \). 
However, at higher mass values, \( \langle v \rangle \) is independent of $\Omega$ but $D_{eff}$ greatly enhances with $\Omega$ as the particles' rotational movement increases.

\section{Discussion}
\label{sec:discussion}

The study of the inertial ABPs in a two-dimensional channel reveals a rich spectrum of transport behaviours arising from the interplay of mass, self-propulsion, and external flow.  
As observed in the overdamped limit, swimmers exhibit oscillatory upstream rheotaxis, characterized by periodic cross-channel motion against the flow, directly reminiscent of \textit{E. coli} and \textit{T. brucei} dynamics~\cite{heddergott2012} and closely aligned with recent observations of droplets in confined flows~\cite{Dey2022}. This agreement validates the essential physics captured by our model and underscores its relevance for real and synthetic microswimmers. Accounting for inertia, the study uncovers qualitatively new regimes. Increasing particle mass weakens upstream motion, suppresses cross-stream oscillations, and promotes wall accumulation.
The mean downstream velocity exhibits a non-monotonic dependence on mass, with an optimal $m_{op}$ that maximizes drift speed; swimmers lighter or heavier than this optimum both drift more slowly. 

To provide an estimate in real units, we consider a swimmer of radius $a \sim 1\ \mu\text{m}$ in water at room temperature ($T \sim 300\ \text{K}$). The translational friction coefficient is $\gamma \sim 2 \times 10^{-3}\ \text{mg/s}$, with a rotational diffusion time $\tau_r \sim 50\ \text{s}$. Typical propulsion speeds are $v_0 \sim 0.2\ \mu\text{m/s}$. For a channel width $y_L = 10\ \mu\text{m}$, the mapping yields an optimal effective mass $m^{op} \sim 0.1\ \text{mg}$. Particles of this mass drift downstream with mean velocities $\sim 0.12\ \mu\text{m/s}$.
These predictions suggest that inertial rheotaxis can be exploited to design lab-on-chip devices for the selective guidance and separation of microswimmers, nanoparticles, and even cellular organelles based on mass.

\section{CONCLUSION}
\label{sec:conclusions}

In this study, we have numerically investigated the transport characteristics of non-interacting inertial active particles in a two-dimensional channel with a Poiseuille flow. Between the channel boundaries, active particles periodically accelerate, decelerate, and continuously change their swimming orientations. The transition from upstream to downstream motion of particles marks a fundamental change in active dynamics, arising when inertial effects become comparable to viscous forces. 
We have shown that inertia fundamentally reshapes microswimmer rheotaxis, e.g., the striking feature of upstream movement of ABPs in the over-damped regime and higher migration at optimal mass. 
We observed that as the mass of ABPs increases, they tend to accumulate near the channel walls and move along the channel boundaries due to the intrinsic nature of the particles and local shear rate due to the applied Poiseuille flow. 
As a result, particle velocity decreases and effective diffusion is greatly enhanced.
These observed findings were further enhanced and controlled by the activity and chirality (rotation rate) of the particles as well as the temperature of the surrounding fluid.  
This establishes inertia as a key control parameter for guiding and separating active particles in confined flows, offering a passive route to mass-selective transport. 
This study provides design principles for microfluidic technologies and invites extension to 3D geometries and hydrodynamic interactions, advancing predictive control of active matter.

\section{ACKNOWLEDGMENT}

This work was financially supported by SERB, Government of India, project-wide CRG/2023/006186.
The authors acknowledge the Indian Institute of Technology (IIT) Kharagpur for providing research facilities.

\bibliographystyle{apsrev4-2}
\bibliography{Draft}

@Article{Gupta2023,
  author    = {Gupta, Ankit and Burada, P. S.},
  journal   = {Physical Review E},
  title     = {Separation of interacting active particles in an asymmetric channel},
  year      = {2023},
  issn      = {2470-0053},
  month     = sep,
  number    = {3},
  pages     = {034605},
  volume    = {108},
  doi       = {10.1103/physreve.108.034605},
  publisher = {American Physical Society (APS)},
}

@Article{Khatri2022,
  author    = {Khatri, Narender and Burada, P. S.},
  journal   = {Physical Review E},
  title     = {Diffusion of chiral active particles in a Poiseuille flow},
  year      = {2022},
  issn      = {2470-0053},
  month     = feb,
  number    = {2},
  pages     = {024604},
  volume    = {105},
  doi       = {10.1103/physreve.105.024604},
  publisher = {American Physical Society (APS)},
}

@Article{Engstler2007,
  author    = {Engstler, Markus and Pfohl, Thomas and Herminghaus, Stephan and Boshart, Michael and Wiegertjes, Geert and Heddergott, Niko and Overath, Peter},
  journal   = {Cell},
  title     = {Hydrodynamic Flow-Mediated Protein Sorting on the Cell Surface of Trypanosomes},
  year      = {2007},
  issn      = {0092-8674},
  month     = nov,
  number    = {3},
  pages     = {505--515},
  volume    = {131},
  doi       = {10.1016/j.cell.2007.08.046},
  publisher = {Elsevier BV},
}

@Article{Galajda2007,
  author    = {Galajda, Peter and Keymer, Juan and Chaikin, Paul and Austin, Robert},
  journal   = {Journal of Bacteriology},
  title     = {A Wall of Funnels Concentrates Swimming Bacteria},
  year      = {2007},
  issn      = {1098-5530},
  month     = dec,
  number    = {23},
  pages     = {8704--8707},
  volume    = {189},
  doi       = {10.1128/jb.01033-07},
  publisher = {American Society for Microbiology},
}

@Article{Rutllant2005,
  author    = {Rutllant, J and López‐Béjar, M and López‐Gatius, F},
  journal   = {Reproduction in Domestic Animals},
  title     = {Ultrastructural and Rheological Properties of Bovine Vaginal Fluid and its Relation to Sperm Motility and Fertilization: a Review},
  year      = {2005},
  issn      = {1439-0531},
  month     = mar,
  number    = {2},
  pages     = {79--86},
  volume    = {40},
  doi       = {10.1111/j.1439-0531.2004.00510.x},
  publisher = {Wiley},
}

@Article{Marchetti2013,
  author    = {Marchetti, M. C. and Joanny, J. F. and Ramaswamy, S. and Liverpool, T. B. and Prost, J. and Rao, Madan and Simha, R. Aditi},
  journal   = {Reviews of Modern Physics},
  title     = {Hydrodynamics of soft active matter},
  year      = {2013},
  issn      = {1539-0756},
  month     = jul,
  number    = {3},
  pages     = {1143--1189},
  volume    = {85},
  doi       = {10.1103/revmodphys.85.1143},
  publisher = {American Physical Society (APS)},
}

@Article{Lauga_2009,
  author    = {Eric Lauga and Thomas R Powers},
  journal   = {Reports on Progress in Physics},
  title     = {The hydrodynamics of swimming microorganisms},
  year      = {2009},
  month     = {aug},
  number    = {9},
  pages     = {096601},
  volume    = {72},
  doi       = {10.1088/0034-4885/72/9/096601},
  publisher = {{IOP} Publishing},
  url       = {https://dx.doi.org/10.1088/0034-4885/72/9/096601},
}

@Article{Bechinger2016,
  author    = {Bechinger, Clemens and Di Leonardo, Roberto and Löwen, Hartmut and Reichhardt, Charles and Volpe, Giorgio and Volpe, Giovanni},
  journal   = {Reviews of Modern Physics},
  title     = {Active Particles in Complex and Crowded Environments},
  year      = {2016},
  issn      = {1539-0756},
  month     = nov,
  number    = {4},
  pages     = {045006},
  volume    = {88},
  doi       = {10.1103/revmodphys.88.045006},
  publisher = {American Physical Society (APS)},
}

@article{Burada_Chemphy,
author = {P. S. Burada and Y. Li and W. Riefler and G. Schmid},
title = {Entropic transport in energetic potentials},
journal = {Chemical Physics},
volume = {375},
number = {2},
pages = {514-517},
year = {2010},
issn = {0301-0104},
doi = {https://doi.org/10.1016/j.chemphys.2010.03.019},
url = {https://www.sciencedirect.com/science/article/pii/S0301010410001126},
}

@Article{Heddergott2012,
  author    = {Heddergott, Niko and Krüger, Timothy and Babu, Sujin B. and Wei, Ai and Stellamanns, Erik and Uppaluri, Sravanti and Pfohl, Thomas and Stark, Holger and Engstler, Markus},
  journal   = {PLoS Pathogens},
  title     = {Trypanosome Motion Represents an Adaptation to the Crowded Environment of the Vertebrate Bloodstream},
  year      = {2012},
  issn      = {1553-7374},
  month     = nov,
  number    = {11},
  pages     = {e1003023},
  volume    = {8},
  url       = {https://journals.plos.org/plospathogens/article?id=10.1371/journal.ppat.1003023},
  editor    = {Beverley, Stephen M.},
  publisher = {Public Library of Science (PLoS)},
}

@Article{Elgeti2015,
  author    = {J Elgeti and R G Winkler and G Gompper},
  journal   = {Reports on Progress in Physics},
  title     = {Physics of microswimmers{\textemdash}single particle motion and collective behavior: a review},
  year      = {2015},
  month     = {apr},
  number    = {5},
  pages     = {056601},
  volume    = {78},
  doi       = {10.1088/0034-4885/78/5/056601},
  publisher = {{IOP} Publishing},
}

@Article{Ramaswamy,
  author    = {Ramaswamy, Sriram},
  journal   = {Annual Review of Condensed Matter Physics},
  title     = {The Mechanics and Statistics of Active Matter},
  year      = {2010},
  month     = {aug},
  number    = {1},
  pages     = {323-345},
  volume    = {1},
  doi       = {10.1146/annurev-conmatphys-070909-104101},
  publisher = {Annual Reviews},
  url       = {https://doi.org/10.1146/annurev-conmatphys-070909-104101},
}

@Article{Ao2014,
  author    = {Ao, X. and Ghosh, P.K. and Li, Y. and Schmid, G. and Hänggi, P. and Marchesoni, F.},
  journal   = {The European Physical Journal Special Topics},
  title     = {Active Brownian motion in a narrow channel},
  year      = {2014},
  issn      = {1951-6401},
  month     = dec,
  number    = {14},
  pages     = {3227--3242},
  volume    = {223},
  doi       = {10.1140/epjst/e2014-02329-1},
  publisher = {Springer Science and Business Media LLC},
}

@Article{Levy2014,
  author    = {Levy, R. and Hill, D. B. and Forest, M. G. and Grotberg, J. B.},
  journal   = {Integrative and Comparative Biology},
  title     = {Pulmonary Fluid Flow Challenges for Experimental and Mathematical Modeling},
  year      = {2014},
  issn      = {1557-7023},
  month     = aug,
  number    = {6},
  pages     = {985--1000},
  volume    = {54},
  doi       = {10.1093/icb/icu107},
  publisher = {Oxford University Press (OUP)},
}

@Article{Bhattacharjee2019,
  author    = {Bhattacharjee, Tapomoy and Datta, Sujit S.},
  journal   = {Nature Communications},
  title     = {Bacterial hopping and trapping in porous media},
  year      = {2019},
  issn      = {2041-1723},
  month     = may,
  number    = {1},
  volume    = {10},
  doi       = {10.1038/s41467-019-10115-1},
  publisher = {Springer Science and Business Media LLC},
}

@Article{Scholz2018,
  author    = {Scholz, Christian and Jahanshahi, Soudeh and Ldov, Anton and Löwen, Hartmut},
  journal   = {Nature Communications},
  title     = {Inertial delay of self-propelled particles},
  year      = {2018},
  issn      = {2041-1723},
  month     = dec,
  number    = {1},
  volume    = {9},
  doi       = {10.1038/s41467-018-07596-x},
  publisher = {Springer Science and Business Media LLC},
}

@Article{Loewen2020,
  author    = {Löwen, Hartmut},
  journal   = {The Journal of Chemical Physics},
  title     = {Inertial effects of self-propelled particles: From active Brownian to active Langevin motion},
  year      = {2020},
  issn      = {1089-7690},
  month     = jan,
  number    = {4},
  volume    = {152},
  doi       = {10.1063/1.5134455},
  publisher = {AIP Publishing},
}

@Article{Wang2012,
  author    = {Wang, S. and Ardekani, A.},
  journal   = {Physics of Fluids},
  title     = {Inertial squirmer},
  year      = {2012},
  issn      = {1089-7666},
  month     = oct,
  number    = {10},
  volume    = {24},
  doi       = {10.1063/1.4758304},
  publisher = {AIP Publishing},
}

@Article{Aghakhani2020,
  author    = {Aghakhani, Amirreza and Yasa, Oncay and Wrede, Paul and Sitti, Metin},
  journal   = {Proceedings of the National Academy of Sciences},
  title     = {Acoustically powered surface-slipping mobile microrobots},
  year      = {2020},
  issn      = {1091-6490},
  month     = feb,
  number    = {7},
  pages     = {3469--3477},
  volume    = {117},
  doi       = {10.1073/pnas.1920099117},
  publisher = {Proceedings of the National Academy of Sciences},
}

@Article{Kim2022,
  author    = {Kim, Yoonho and Zhao, Xuanhe},
  journal   = {Chemical Reviews},
  title     = {Magnetic Soft Materials and Robots},
  year      = {2022},
  issn      = {1520-6890},
  month     = feb,
  number    = {5},
  pages     = {5317--5364},
  volume    = {122},
  doi       = {10.1021/acs.chemrev.1c00481},
  publisher = {American Chemical Society (ACS)},
}

@Article{Zoettl2012,
  author    = {Zöttl, Andreas and Stark, Holger},
  journal   = {Physical Review Letters},
  title     = {Nonlinear Dynamics of a Microswimmer in Poiseuille Flow},
  year      = {2012},
  issn      = {1079-7114},
  month     = may,
  number    = {21},
  pages     = {218104},
  volume    = {108},
  doi       = {10.1103/physrevlett.108.218104},
  publisher = {American Physical Society (APS)},
}

@Article{Zoettl2013,
  author    = {Zöttl, Andreas and Stark, Holger},
  journal   = {The European Physical Journal E},
  title     = {Periodic and quasiperiodic motion of an elongated microswimmer in Poiseuille flow},
  year      = {2013},
  issn      = {1292-895X},
  month     = jan,
  number    = {1},
  volume    = {36},
  doi       = {10.1140/epje/i2013-13004-5},
  publisher = {Springer Science and Business Media LLC},
}

@Article{Ren2017,
  author    = {Ren, Liqiang and Zhou, Dekai and Mao, Zhangming and Xu, Pengtao and Huang, Tony Jun and Mallouk, Thomas E.},
  journal   = {ACS Nano},
  title     = {Rheotaxis of Bimetallic Micromotors Driven by Chemical–Acoustic Hybrid Power},
  year      = {2017},
  issn      = {1936-086X},
  month     = sep,
  number    = {10},
  pages     = {10591--10598},
  volume    = {11},
  doi       = {10.1021/acsnano.7b06107},
  publisher = {American Chemical Society (ACS)},
}

@Article{Dey2022,
  author    = {Dey, Ranabir and Buness, Carola M. and Hokmabad, Babak Vajdi and Jin, Chenyu and Maass, Corinna C.},
  journal   = {Nature Communications},
  title     = {Oscillatory rheotaxis of artificial swimmers in microchannels},
  year      = {2022},
  issn      = {2041-1723},
  month     = may,
  number    = {1},
  volume    = {13},
  doi       = {10.1038/s41467-022-30611-1},
  publisher = {Springer Science and Business Media LLC},
}

@Article{Rusconi2014,
  author    = {Rusconi, Roberto and Guasto, Jeffrey S. and Stocker, Roman},
  journal   = {Nature Physics},
  title     = {Bacterial transport suppressed by fluid shear},
  year      = {2014},
  issn      = {1745-2481},
  month     = feb,
  number    = {3},
  pages     = {212--217},
  volume    = {10},
  doi       = {10.1038/nphys2883},
  publisher = {Springer Science and Business Media LLC},
}

@Article{Maity2022,
  author    = {Maity, Ruma and Burada, P.S.},
  journal   = {Journal of Fluid Mechanics},
  title     = {Unsteady chiral swimmer and its response to a chemical gradient},
  year      = {2022},
  issn      = {1469-7645},
  month     = apr,
  volume    = {940},
  doi       = {10.1017/jfm.2022.239},
  publisher = {Cambridge University Press (CUP)},
}

@article{Suresh,
title = {Biomechanics and biophysics of cancer cells},
journal = {Acta Materialia},
volume = {55},
number = {12},
pages = {3989-4014},
year = {2007},
issn = {1359-6454},
doi = {https://doi.org/10.1016/j.actamat.2007.04.022},
url = {https://www.sciencedirect.com/science/article/pii/S1359645407002789},
author = {Subra Suresh}
}

@Article{Ma2015,
  author    = {Ma, Xing and Hahn, Kersten and Sanchez, Samuel},
  journal   = {Journal of the American Chemical Society},
  title     = {Catalytic Mesoporous Janus Nanomotors for Active Cargo Delivery},
  year      = {2015},
  issn      = {1520-5126},
  month     = apr,
  number    = {15},
  pages     = {4976--4979},
  volume    = {137},
  doi       = {10.1021/jacs.5b02700},
  publisher = {American Chemical Society (ACS)},
}

@Article{Soler2014,
  author    = {Soler, Lluís and Sánchez, Samuel},
  journal   = {Nanoscale},
  title     = {Catalytic nanomotors for environmental monitoring and water remediation},
  year      = {2014},
  issn      = {2040-3372},
  number    = {13},
  pages     = {7175--7182},
  volume    = {6},
  doi       = {10.1039/c4nr01321b},
  publisher = {Royal Society of Chemistry (RSC)},
}

@Article{Gao2014,
  author    = {Gao, Wei and Wang, Joseph},
  journal   = {ACS Nano},
  title     = {The Environmental Impact of Micro/Nanomachines: A Review},
  year      = {2014},
  issn      = {1936-086X},
  month     = mar,
  number    = {4},
  pages     = {3170--3180},
  volume    = {8},
  doi       = {10.1021/nn500077a},
  publisher = {American Chemical Society (ACS)},
}

@Article{Peng2020,
  author    = {Peng, Zhiwei and Brady, John F.},
  journal   = {Physical Review Fluids},
  title     = {Upstream swimming and Taylor dispersion of active Brownian particles},
  year      = {2020},
  issn      = {2469-990X},
  month     = jul,
  number    = {7},
  pages     = {073102},
  volume    = {5},
  doi       = {10.1103/physrevfluids.5.073102},
  publisher = {American Physical Society (APS)},
}

@Article{Reguera2012,
  author    = {Reguera, D. and Luque, A. and Burada, P. S. and Schmid, G. and Rub\'{\i}, J. M. and H\"anggi, P.},
  journal   = {Phys. Rev. Lett.},
  title     = {Entropic Splitter for Particle Separation},
  year      = {2012},
  issn      = {1079-7114},
  month     = {Jan},
  number    = {2},
  pages     = {020604},
  volume    = {108},
  doi       = {10.1103/PhysRevLett.108.020604},
  issue     = {2},
  numpages  = {5},
  publisher = {American Physical Society},
  url       = {https://link.aps.org/doi/10.1103/PhysRevLett.108.020604},
}

@Article{Jin2014,
  author    = {Jin, Chao and McFaul, Sarah M. and Duffy, Simon P. and Deng, Xiaoyan and Tavassoli, Peyman and Black, Peter C. and Ma, Hongshen},
  journal   = {Lab Chip},
  title     = {Technologies for label-free separation of circulating tumor cells: from historical foundations to recent developments},
  year      = {2014},
  issn      = {1473-0189},
  number    = {1},
  pages     = {32--44},
  volume    = {14},
  doi       = {10.1039/c3lc50625h},
  publisher = {Royal Society of Chemistry (RSC)},
}

@Article{Elgeti2013,
  author    = {Elgeti, Jens and Gompper, Gerhard},
  journal   = {EPL (Europhysics Letters)},
  title     = {Wall accumulation of self-propelled spheres},
  year      = {2013},
  issn      = {1286-4854},
  month     = feb,
  number    = {4},
  pages     = {48003},
  volume    = {101},
  doi       = {10.1209/0295-5075/101/48003},
  publisher = {IOP Publishing},
}

@Article{VillalobosConcha2025,
  author    = {Villalobos-Concha, Cristian and Liu, Zhengyang and Ramos, Gabriel and Goral, Martyna and Lindner, Anke and López-León, Teresa and Clément, Eric and Soto, Rodrigo and Cordero, María Luisa},
  journal   = {Proceedings of the National Academy of Sciences},
  title     = {Active bacterial baths in droplets},
  year      = {2025},
  issn      = {1091-6490},
  month     = jul,
  number    = {31},
  volume    = {122},
  doi       = {10.1073/pnas.2426096122},
  publisher = {Proceedings of the National Academy of Sciences},
}

@Article{SchwarzLinek2012,
  author    = {Schwarz-Linek, J. and Valeriani, C. and Cacciuto, A. and Cates, M. E. and Marenduzzo, D. and Morozov, A. N. and Poon, W. C. K.},
  journal   = {Proceedings of the National Academy of Sciences},
  title     = {Phase separation and rotor self-assembly in active particle suspensions},
  year      = {2012},
  issn      = {1091-6490},
  month     = mar,
  number    = {11},
  pages     = {4052--4057},
  volume    = {109},
  doi       = {10.1073/pnas.1116334109},
  publisher = {Proceedings of the National Academy of Sciences},
}

@Article{Kesteren2023,
  author    = {van Kesteren, Steven and Alvarez, Laura and Arrese-Igor, Silvia and Alegria, Angel and Isa, Lucio},
  journal   = {Proceedings of the National Academy of Sciences},
  title     = {Self-propelling colloids with finite state dynamics},
  year      = {2023},
  issn      = {1091-6490},
  month     = mar,
  number    = {11},
  volume    = {120},
  doi       = {10.1073/pnas.2213481120},
  publisher = {Proceedings of the National Academy of Sciences},
}

@Article{Palacci2014,
  author    = {Palacci, J. and Sacanna, S. and Kim, S.-H. and Yi, G.-R. and Pine, D. J. and Chaikin, P. M.},
  journal   = {Philosophical Transactions of the Royal Society A: Mathematical, Physical and Engineering Sciences},
  title     = {Light-activated self-propelled colloids},
  year      = {2014},
  issn      = {1471-2962},
  month     = nov,
  number    = {2029},
  pages     = {20130372},
  volume    = {372},
  doi       = {10.1098/rsta.2013.0372},
  publisher = {The Royal Society},
}

\end{document}